\documentclass[12pt, prd, showpacs]{revtex4}

\usepackage{amsmath}
\usepackage{amssymb}

\begin{document}

\title{Quasi-black holes: definition and general properties}
\author{Jos\'{e} P. S. Lemos}
\affiliation{Centro Multidisciplinar de Astrof\'{\i}sica - CENTRA,
Departamento de F\'{\i}%
sica, Instituto Superior T\'ecnico - IST, Universidade T\'{e}cnica de Lisboa
- UTL, Av. Rovisco Pais 1, 1049-001 Lisboa, Portugal\,\,}
\email{lemos@fisica.ist.utl.pt}
\author{Oleg B. Zaslavskii}
\affiliation{Astronomical Institute of Kharkov V.N. Karazin National
University, 35
Sumskaya St., Kharkov, 61022, Ukraine}
\email{ozaslav@kharkov.ua}

\begin{abstract}
Objects that are on the verge of being extremal black holes but
actually are distinct in many ways are called quasi-black
holes. Quasi-black holes are defined here and treated in a unified way
through the displaying of their properties. The main ones are (i)
there are infinite redshift whole regions, (ii) the spacetimes exhibit
degenerate, almost singular, features but their curvature invariants
remain perfectly regular everywhere, (iii) in the limit under
discussion, outer and inner regions become mutually impenetrable and
disjoint, although, in contrast to the usual black holes, this
separation is of a dynamical nature, rather than purely causal, (iv)
for external far away observers the spacetime is virtually
indistinguishable from that of extremal black holes. It is shown, in
addition, that quasi-black holes must be extremal. Connections with
black hole and wormhole physics are also drawn.
\end{abstract}

\maketitle



\newpage 

\section{Introduction}

A quasi-black hole (QBH) is neither a usual regular spacetime, as for
instance a star, nor a black hole (BH). So what is it? Roughly speaking one
can say that a QBH is an object on the verge of becoming an extremal BH but
actually is distinct from it in many ways. In the present paper we show
that, among other properties, perhaps the most striking ones that
characterize a QBH are that there is an infinite redshift region, the
spacetime exhibits degenerate features with the curvature invariants
remaining perfectly regular everywhere, outer and inner regions become
mutually impenetrable and disjoint, and for an external observer at infinity
the spacetime is indistinguishable from that of an extremal black hole.
Another interesting feature is that a QBH has to be extremal.

To try to understand how a QBH may arise, we note that, remarkably, contrary
to the common case where instabilities set in much before a matter system
reaches its own gravitational radius, there are some systems for which the
gravitational radius can be approached in a sequence of static
configurations. They were first noticed within the context of
Majumdar-Papapetrou systems \cite{maj,pt} by Bonnor in 
\cite{bonnor1964,bonnor1975,bonnor1999} and are systems composed of
extremal charged dust, where the energy density is equal to the charge
density, with no pressure term, and joined to an asymptotically flat
extremal Reissner-Nordstr\"{o}m region. These systems are called
Bonnor stars. In recent years, the interest in such objects was
renewed due to further investigations on their properties, where it
was found that they have very interesting properties such as the
formation of a QBH state \cite{kleberlemoszanchin,jmp}. The same set
of properties had also been found in extended Bonnor stars
\cite{exdust}, extremal systems with a more sophisticated density
distribution. One can then say that QBHs can be thought of as the end
state of a sequence, of gradually more compact, quasistatic
appropriate Majumdar-Papapetrou configurations, such as Bonnor stars
and extended Bonnor stars. The most surprising feature of all these
systems is the fact that the limiting case, at the threshold of the
formation of an event horizon, is very peculiar. Although to external
observers the system looks like an extremal black hole, its internal
properties, so to speak, are very different from what one could expect
in the case of a usual BH. In the limit, instead of an extremal BH one
has a QBH, and instead of an event horizon one has a
quasihorizon. This then expands the existing taxonomy of relativistic
objects, adding to it something that is neither a usual regular
spacetime, a star, nor a BH, it is a QBH. There are other systems that
display QBHs. In self-gravitating Higgs magnetic monopole systems, a
seemingly different system, it was also found, in a totally
independent way, that in a certain well defined limit a QBH appears as
a natural state, and it was indeed within these studies that the term
QBH was coined \cite{lw1,lw2}. The similarity of the properties of the
Bonnor stars and gravitational magnetic monopoles was clearly
recognized in \cite{jmp}. Both kinds of systems look quite
physical. For example, the Bonnor star system can be realized when a
sphere of neutral hydrogen has lost a fraction 10$^{-18}$ of its
electrons, while magnetic monopoles should be formed if standard grand
unified theories prove to be correct. In addition, and surprisingly,
similar objects with QBH properties, were found for composite
spacetimes even in the case of pure electrovacuum \cite{mod} (see also
\cite{n}). These vacuum systems are composed of an exterior
Reissner-Nordstr\"{o}m part glued to an inner Bertotti-Robinson
spacetime (see \cite{br1,br2,lap,geroch,rn-br1,rn-br2}), or of an
exterior Reissner-Nordstr\"{o}m part glued to an an inner Minkowski
spacetime \cite{vf}. It is interesting to note that in a certain
sense some of these systems realize the idea of charge without charge
\cite{wheeler1}.

There are at least eight related subjects connected to QBHs. Not all
of them will be analyzed in detail, since that would take us far a
field. The first connection is with naked BHs \cite{nk1,nk2,vo,ssn}
(see also \cite{bm} for the issue of the behavior of the quasilocal
energy and momentum under boosts from a static frame to a free-falling
one). Naked BHs have diverging Riemann tensors in certain physically
meaningful frames, which in turn relates them to the singular or
regular character of a spacetime and the cosmic censorship hypothesis
\cite{es}. As we will see, QBHs have this naked property. The second
connection is with the end state of an extremal matter
configuration. One can ask whether a QBH can be attained
physically. Can a QBH configuration be reached through a finite number
of steps from a regular configuration? This is related to whether an
analogue of the third law of BH thermodynamics (see, e.g.,
\cite{wald}) is enforced or not for QBHs. The third connection is with
the instabilities that might set in before the gravitational radius is
reached. This is related to the Buchdahl limit \cite{buchdahl}, i.e.,
the minimum radius to mass ratio $r_0/m$ that a stable configuration
can have, where $r_0$ and $m$ are the radius and mass of the
configuration, respectively.  For perfect fluid spheres it is
$r_0/m\geq 9/4$, while for charged spheres the ratio decreases, it
goes to $r_0/m\geq 1$ precisely in the case of extremal charged dust
\cite{df,simingyunqiang}. The fourth connection is with the hoop
conjecture \cite{thorne}, since it seems that QBHs grossly violate
it. The conjecture states that a BH forms when matter of mass $M$ is
compacted within a given definite hoop, in \cite{thorne} taken to be
$\sim 4\pi M$ ($G=C=1$), and shown later in \cite{bonnor} that the
hoop should be reduced for extremal charged matter to $\sim 2\pi
M$. But as it will be shown, for extremal matter a BH never arises,
instead a QBH forms in the limit. The fifth connection is with the no
hair theorems. It was conjectured by Wheeler that BHs should have no
hair, in particular no electromagnetic hair, a conjecture that has
been verified \cite {wheeler2,bekenstein}. On the other hand QBHs have
the feature that they may have some electromagnetic hair
\cite{exdust}, adding to the list of distinct properties between both
objects. The sixth connection is with Bardeen BHs
\cite{bardeen,beato}, i.e., BHs that have a kind of magnetic charged
matter inside the horizon, and have no singularities inside. Following
a theorem by Borde \cite{borde} this means the topology inside the
horizon is different from the usual one. Now, the configurations we
are studying are neither usual BHs nor Bardeen BHs, they are
QBHs. They have quasihorizons and the Kretschmann scalar is finite
inside, although, as we will see, this does not exclude other
degenerate features. So, it appears that in order to avoid a true
horizon without a singularity inside with a consequent change of
topology, the object opts to form a QBH, instead of a Bardeen-type
BH. The seventh connection is with objects that mimic BHs. For
instance, wormholes (see, e.g., \cite{bookvisser}) can be good
mimickers of BH properties \cite {damoursoloduk}. Although, QBHs and
BHs share many properties from the viewpoint of an external observer,
the full study of this subject has not been done. The eighth
connection is with the entropy issue. For the usual BHs one does not
yet know for sure where are the degrees of freedom and thus how their
entropy arises (see, e.g., \cite{lemosreview}). For QBHs it seems that
the entropy comes from the entangled fields hidden beyond the quasi
horizon \cite{wl}. There are possible connections with other subjects,
like gravitational collapse (which has not been studied for the case
of extremal matter) and vacuum polarization effects, to name two.

In this work we obtain and analyze the geometric and physical properties of
a QBH. The paper is organized as follows. In section \ref{def} a definition
of a QBH is given. In section \ref{prop} the properties of QBHs are
displayed in several instances. Initially, we study Bonnor stars, both
truncated and extended, then we analyze gravitational magnetic monopoles,
and finally we study glued extremal vacua. All these instances of QBHs show
a number of similar properties. In section \ref{furtherprop} we prove an
important theorem that states that QBHs have to be extremal. In section V we
discuss the relationship between regular and singular features in QBHs
spacetime and whether the QBH state can be physically attained. Finally in
section \ref{conc} we draw some interesting conclusions.

\section{Definition of quasi black holes}

\label{def}

The fact that so different kinds of physical systems like extremal dust,
Yang-Mills$-$Higgs matter, and composite vacuum systems, may exhibit the
same qualitative features suggests that the unusual properties of QBHs can
be explained in an unified manner. So first we define what a QBH is, and
then we investigate in detail the properties of such a system in the various
instances.

A QBH can be defined as an object with the following properties. Consider
the static spherically symmetric metric, often written as
\begin{equation}
ds^{2}=-B(r)\,dt^{2}+A(r)\,dr^{2}+r^{2}\,d\Omega ^{2}\,,  \label{ab}
\end{equation}
where $r$ is the Schwarzschild radial coordinate, $d\Omega ^{2}=d\theta
^{2}+\sin ^{2}\theta \,d\phi ^{2}$, and $B(r)$ and $A(r)$ are metric
potentials. It is useful to define a new metric potential $V$ through
\begin{equation}
V(r)=\frac{1}{A(r)}\,.  \label{defintionofV}
\end{equation}
Let an inner matter configuration, with an asymptotic flat exterior region,
exist with the properties (a) the function $V(r)$ attains a minimum at some
$r^{\ast }\neq 0$, such that $V(r^{\ast })=\varepsilon $, with $\varepsilon
<<1$, this minimum being achieved either from both sides of $r^{\ast }$ or
from $r>r^{\ast }$ alone, (b) for such a small but nonzero $\varepsilon $
the configuration is regular everywhere with a nonvanishing metric function
$B$, at most the metric contains only delta-function like shells, and (c) in
the limit $\varepsilon \rightarrow 0$ the metric coefficient $B\rightarrow
0$ for all $r\leq r^{\ast }$. These three features define a QBH. Note that
although the above definition of QBHs relies on the coordinate system and
metric coefficient $V$ given in equations (\ref{ab})-(\ref{defintionofV}),
actually, this definition can be done in a form invariant under the choice
of the radial coordinate. Indeed, it is sufficient to replace $V$ by $
(\nabla r)^{2}$. In the Schwarzschild coordinates of equation (\ref{ab}) one
has $(\nabla r)^{2}=V$.

In turn, these three features entail some nontrivial consequences: (i) there
are infinite redshift whole regions, (ii) when $\varepsilon \rightarrow 0$,
a free-falling observer finds in his own frame infinitely large tidal forces
in the whole inner region, showing some form of degeneracy, although the
spacetime curvature invariants remain perfectly regular everywhere, (iii) in
the limit, outer and inner regions become mutually impenetrable and
disjoint, and on can also show that (iv) for external far away observers the
spacetime is virtually indistinguishable from that of extremal black holes.
In addition, QBHs must be extremal. The QBH is on the verge of forming an
event horizon, but it never forms one, instead, a quasihorizon appears. For
a QBH the metric is well defined and everywhere regular. However,
properties, such as when $\varepsilon=0$, QBH spacetimes become degenerate,
almost singular, have to be examined with care.

\section{Properties of quasi-black holes}

\label{prop}

Now, we study the three different examples of QBH behavior separately
(namely, extremal charged dust, Yang-Mills$-$Higgs matter, and composite
vacuum systems), to show how the same features reveal themselves in these
different circumstances.

\subsection{Extremal charged dust and Bonnor stars}

\label{extremaldustbonnor} Within extremal charged dust there are two
different cases worth of study, namely the ones studied by Bonnor \cite
{bonnor1964,bonnor1975,bonnor1999,kleberlemoszanchin,jmp} and the ones
studied by Lemos and Weinberg \cite{exdust}, both systems belong to the
Majumdar-Papapetrou class \cite{maj,pt}.

\subsubsection{Bonnor stars: bounded distribution of extremal dust matched
to an electrovacuum at $r=r_{0}$ (with $r_0>m$) \protect\cite
{bonnor1964,bonnor1975,bonnor1999,kleberlemoszanchin,jmp}}

\textit{Generic properties:}

The radius $r_{0}$ is the boundary of the star. Inside there is matter,
outside there is vacuum. So, in the region $r>r_{0}$ the metric potential
$V$ has the usual form for the extremal Reissner-Nordstr\"{o}m BH,
\begin{equation}
V^{\mathrm{RN}}=\left( 1-\frac{m}{r}\right) ^{2}\,,
\label{reissnernordstrom-b}
\end{equation}
where $m$ is the total mass. For $r\leq r_{0}$ it is described by a
Majumdar-Papapetrou type solution which in Schwarzschild-like coordinates
can be written as
\begin{equation}
V=\left( 1-\frac{\mu (r)}{r}\right) ^{2}\,,  \label{vm}
\end{equation}
with, the mass density $\rho $ and the function $\mu $ being connected
through
\begin{equation}
4\pi \rho =\frac{\mu ^{\prime }}{r^{2}}\left( 1-\frac{\mu }{r}\right) \,.
\label{rom}
\end{equation}
The function $\mu(r)$ can be interpreted as the proper mass enclosed within
a sphere of a radius $r$. Similarly, we define $e(r)$ as the proper charge
enclosed within a sphere of a radius $r$. For Majumdar-Papapetrou systems 
$\mu(r)=e(r)$. We want to glue smoothly both regions, so $\mu(r_{0})=m$ and
for $r\leq r_{0}$,
\begin{equation}
\sqrt{B(r)}=\sqrt{B^{\mathrm{RN}}(r_{0})}\,\exp (\nu )\,,  \label{bm}
\end{equation}
where,
\begin{equation}
\nu =\int_{r_{0}}^{r}dr\,\frac{\mu }{r^{2}\,(1-\frac{\mu }{r})}.
\label{nufunction}
\end{equation}
This guarantees that on the boundary, $\sqrt{B(r_{0}-0)}=\sqrt{B(r_{0}+0)}$.

\vskip 0.5cm \textit{Proper spatial distance:}

The proper distance can be written as
\begin{equation}
l=\int dr\,\frac{1}{\sqrt{V(r)}}=\int dr\,
\frac{1}{\left( 1-\frac{\mu }{r}
\right) }=\int dr\,\frac{\mu ^{\prime }}{4\pi \rho r^{2}}\,.  
\label{proper}
\end{equation}
If $\rho $ remains finite and nonzero in the quasihorizon limit
$r_{0}\rightarrow m$, like in the special examples of \cite
{bonnor1964,bonnor1975,bonnor1999,kleberlemoszanchin,jmp}, one can
obtain from (\ref{rom}) that $\mu \approx m-[8\pi \rho
(m)m^{3}]^{1/2}(r-m)^{-1/2}$ near $r=m$ and, thus, the integral
(\ref{proper}) converges and the proper distance from any interior
point to the boundary, from the inside, remains finite accordingly. In
terms of the potential $V(r)$ we can see this by noting that when
$V(r)$ has a single root the integral in (\ref{proper}) is finite,
when it has a double root the integral behaves logarithmically and
yields an infinite result. So, from the inside one has $\mathrm{lim}
_{r_{0}\rightarrow m}^{r\rightarrow r_{0}}V^{\prime }(r)=-8\pi \rho
(r_{0})\,r_{0}|_{r_{0}=m}<0$, the root is simple and the proper
distance is finite. On the other hand, from the outside, $V^{\prime
}(r)|_{r_{0}\rightarrow m}\rightarrow 0$, one has thus a double root
in the limit, and the proper distance is infinite, yielding a
semi-infinite throat from the outside, which is a well known result
for the extremal Reissner-Nordstr\"{o}m geometry.

\vskip0.5cm \textit{Motion of massive and massless particles:}

 From the inside to the outside, the existence of an impenetrable
barrier: Now, let us consider the motion of particles in this
spacetime. In doing so, an important question concerns the transition
from the inner region to the outer one. In the interior the suitable
time variable measured by a static observer can be obtained by
rescaling the time $t$, such that $t=\frac{
\tilde{t}}{\sqrt{B^{\mathrm{RN}}(r_{0})}}$. So $\sqrt{\tilde{B}}\equiv
\frac{ \sqrt{B}}{\sqrt{B^{\mathrm{RN}}(r_{0})}}$ is finite. Now, if a
timelike particle is emitted in the radial upward direction with the
finite energy $ \tilde{E}$ one can easily find from the conservation
law that the proper time $\tilde{\tau}$ is equal to $\tilde{\tau}=\int
dl\frac{\sqrt{\tilde{B}}}{
\sqrt{\tilde{E}^{2}-{\tilde{B}}}}\,,\label{tp}$ where, with respect to
time $\tilde{t}$, we have put the proper mass of a timelike particle
equal to one. Thus, the particle reaches the border in a finite proper
time $\tilde{ \tau}$. The quantity $\sqrt{\tilde{B}}$ is finite
everywhere inside and is equal to unity on the border. However,
outside one has that $\sqrt{\tilde{B}} =\frac{(r-m)r_{0}}{(r_{0}-m)r}$
grows without bound when one takes the limit $r_{0}\rightarrow m$ for
any $r>m$. When denominator in the equation above vanishes, it gives a
turning point at $r_{1}=\frac{mr_{0}}{r_{0}-\tilde{E} (r_{0}-m)}$. In
the limit $r_{0}\rightarrow m$ we have also $ r_{1}\rightarrow m$ for
any finite $\tilde{E}$. Thus, the boundary between the matter and
vacuum regions, acts like an infinite barrier which prevents particles
from penetrating into the outer region from inside. For zero rest mass
particles, like photons, moving radially, the affine parameter
$\lambda $ is given by $\lambda =\tilde{\omega}^{-1}\int
dl\sqrt{\tilde{B}}$, where $\tilde{\omega}$ is the photon frequency
measured with respect to the time $ \tilde{t}$. Then in the outer
region one has $\lambda (r)-\lambda (r_{0})=
\frac{(r-m)r_{0}}{(r_{0}-m)r}$. This difference becomes infinite in
the limit $r_{0}\rightarrow m$ for any $r>m$. As a result, again the
boundary in the limit under discussion acts as a impenetrable
barrier. Thus it also acts as a lightlike infinity.

 From the outside to the inside, shrinking interval of proper time,
tidal forces and naked behavior: (i) Shrinking interval of proper 
time - it follows from the above formulas that in the limit
$B^{\mathrm{RN} }(r_{0})\rightarrow 0$ the finite interval in time
$\tilde{t}$ correspond to infinitely delayed intervals in time
$t$. However, if one calculated the proper time for an infalling
particle moving with the energy $E$ from the outside (which is defined
with respect to the time at infinity, without rescaling) it follows
from $\tau =\int dl\frac{\sqrt{B}}{\sqrt{E^{2}-{B}}}$ that $\tau
\rightarrow 0$ between any two points inside since $B\rightarrow 0 $
there, while the proper distance is finite for bounded Bonnor stars as
is explained after eq. (\ref{proper}). Manifestations of these general
properties for self-gravitating monopole spacetimes were discussed in
\cite {lw2}. (ii) Tidal forces and naked behavior - to understand the
existence of naked behavior for these systems we have to compute the
Riemann tensor in a freely falling frame. First we compute it in the
static coordinate frame.  Consider then the behavior of the
nonvanishing Riemann tensor components.  One has, $\quad
R_{\hat{0}\hat{r}}^{\hat{0}\hat{r} }\equiv K=-V\frac{\sqrt{B}
^{\,\prime \prime }}{\sqrt{B}}-\frac{V^{\prime }}{ 2 }\frac{\sqrt{B}
^{\,\prime }}{\sqrt{B}}\,,$ $\quad R_{\hat{0}\hat{\theta} }^{
\hat{0}\hat{ \theta}}=\bar{K}(r)=-\frac{V}{r}\frac{\sqrt{B}^{\,\prime
}}{ \sqrt{B}}\,,$ $\quad
R_{\hat{\phi}\hat{\theta}}^{\hat{\phi}\hat{\theta} }\equiv F(r)=\frac{
1 }{r^{2}}(1-V)\,,$ $\quad R_{\hat{\theta}\hat{r}}^{\hat{
\theta}\hat{r} }\equiv \bar{F}(r)=-\frac{V^{\prime }}{2r}\,$. One can
then obtain directly that all these components remain finite in the
inner region in the limit $ \sqrt{B(r_{0})}\rightarrow 0$. Indeed, it
follows from (\ref{vm})-(\ref{bm}) that the quantities defined above
are given by, $K=\frac{ 2\mu }{r^{3}}-3 \frac{\mu ^{2}}{r^{4}}\,,$
$\bar{K}(r)=-\frac{\mu }{r^{3}}\,,$ $F=\frac{1}{ r^{2}}\left(
2\frac{\mu }{f}-\frac{\mu ^{2}}{r^{2}}\right) \,,$ $\bar{F} (r)=-8\pi
r\rho \,.$ So for finite $\rho $ and $\mu $, the above quantities are
obviously finite everywhere in the inner region including the boundary
and origin. Correspondingly, the Kretschmann scalar is finite and the
geometry is regular in spite of the fact that the metric function
$\sqrt{B}$, suited to the time variable of an asymptotically flat
observer, vanishes everywhere in the inner region. Having computed the
Riemann tensor in the static coordinate frame we can now go on to a
free-falling frame. Here, the situation becomes more subtle. We have
now enhancement of the curvature components. To see this, write first,
$Z\equiv (\bar{F}-\bar{K})$. Then, $
\tilde{Z}=Z\,(2\frac{E^{2}}{B}-1)$, where $E$ is the energy of the
freely falling particle, representing the freely falling particle
frame. So, one sees that in the limit $\sqrt{B}\rightarrow 0$, these
components of the curvature tensor and the corresponding tidal forces
grow without bound.  Thus, we encounter behavior typical of naked BHs
\cite{nk1,nk2} (see also \cite{vo,ssn,bm}), although in the present
case we have QBHs instead of BHs.  Note, in passing, that naked
behavior is consistent with the regularity of the geometry in the
static frame since in the free-falling frame different terms enter the
expression in the Kretschmann scalar with different signs and may
mutually cancel.

\vskip 0.5cm \textit{Redshift:}

Bonnor stars, in the limit of QBH formation, display infinite redshift
phenomenon as shown in 
\cite{bonnor1964,bonnor1975,bonnor1999,kleberlemoszanchin,jmp}, where
it is assumed that the frequency is measured with respect to time $t$
at infinity.  However, we have seen that two scales of time, $t$ and
$\tilde t$, are relevant for the systems under discussion.

In terms of the time $t$, the outer time, the product $\omega \sqrt{B(r)}
=\omega _{\mathrm{c}}$ remains constant on each ray during the propagation
of light in a static gravitational field, where here $\omega _{\mathrm{c}}$
is some constant frequency. Note that $\omega _{\mathrm{c}}$ is a frequency
measured with respect to time $t$, $\omega $ is the frequency measured with
respect to the proper time at a given point $r$. Since at infinity $B=1$,
one obtains that $\omega _{\mathrm{c}}$ is $\omega (r\rightarrow \infty
)\equiv \omega _{\infty }$, so that one can write $\omega \sqrt{B(r)}=\omega
_{\infty }$. A distant observer would register an infinite redshift ($\omega
_{\mathrm{c}}\rightarrow 0$) if an emitted particle had a finite $\omega $
inside the matter since $B\rightarrow 0$ there in the QBH limit. Only
high-frequency photons with infinite $\omega $ inside the quasihorizon but
finite $\omega _{\infty }$ can escape to infinity. This occurs for any
Bonnor star whose boundary gets arbitrarily close to the horizon ($B^{
\mathrm{RN}}(r_{0})\rightarrow 0$), this property being model-independent.

In terms of the time $\tilde{t}$, the inner time, an observer uses
rather the equality $\omega
\sqrt{\tilde{B}}=\tilde{\omega}_{\mathrm{c}}$, where $
\sqrt{\tilde{B}}=\frac{B}{\sqrt{B^{\mathrm{RN}}(r_{0})}}$, as above,
and $\tilde{\omega}_{\mathrm{c}}=
\frac{\omega_\infty}{\sqrt{B^{\mathrm{RN}}(r_{0}
) }}$. The observer does not encounter an infinitely large redshift
since in the inner region $\tilde{\omega}_{\mathrm{c}}$ and
$\tilde{B}$ remain finite and nonzero, even in the QBH limit when
$B^{\mathrm{RN}}(r_{0})\rightarrow0$. However, we have seen before
that the latter property causes an infinite barrier for particles
moving outward.

Thus, both properties (infinite redshift for an inner signal, emitted inside
and registered by an observer at infinity, and impenetrable barrier for
particles moving from the inner to the outer region) are different
consequences of the same property $B^{\mathrm{RN}}(r_{0})\rightarrow0$.

\vskip0.5cm \textit{Other considerations: the end state}

The QBH can be considered as the end state of a sequence of ever more
compact Bonnor stars. There is no way in which one can get a more compact
object from it, or somehow turn it into an extremal BH. Whether this end
state can be achieved by a physical process is a thorny issue that will be
discussed towards the end of this article.

\vskip0.5cm \textit{Example:}

We demonstrate now, using an explicit example of a Bonnor star given in
\cite
{bonnor1999}, what happens to the metric in the quasihorizon limit. The
metric of any spherically-symmetrical Majumdar-Papapetrou system can be
written in isotropic coordinates as (see \cite{maj,pt}),
\begin{equation}
ds^{2}=-B\,dt^{2}+B^{-1}\,\left( dR^{2}+R^{2}d\Omega ^{2}\right) \,
\label{dust0}
\end{equation}
where the radial coordinate $R$ is related to the Schwarzschild coordinate
$r$ of equation (\ref{ab}) by
\begin{equation}
R=r\,\sqrt{B}\text{.}  \label{isoR}
\end{equation}
 From \cite{bonnor1999}, defining a new potential $U(R)$ as
$U=1/\sqrt{B}$, a good choice for the internal and external $U$,
$U^{I}$ and $U^{E}$ respectively, is
\begin{equation}
U^{I}=1+\frac{m}{R_{0}}+\frac{m(R_{0}^{n}-R^{n})}{nR_{0}^{n+1}}\,,\quad
0\leq R\leq R_{0}\text{,}  \label{uint}
\end{equation}
\begin{equation}
U^{E}=1+\frac{m}{R}\,,\quad R\geq R_{0}>0\,,  \label{uext}
\end{equation}
where $m$ is the mass of the configuration, $R_{0}$ is the boundary of
the star, and $n$ is a free exponent, with $n\geq 2$, $n=2$ being a
typical case. The extremal charged dust occupies the region $0\leq
R\leq R_{0}$. For $R>R_{0}$ the metric represents an external extremal
Reissner-Nordstr\"{o}m metric. In this outer region the relation
between $r$ and $R$ is simple, $ r=R+m$. Then the boundary areal
radius $r_{0}$ is given by $r_{0}=m+R_{0}$.  When $R_{0}\rightarrow 0$
the areal radius $r_{0}$ of the boundary approaches that of the
quasihorizon as closely as one likes, with the dust density remaining
finite everywhere inside, including the boundary. Let us take the next
step to obtain the limiting metric explicitly. It is convenient to
make the following substitutions for the interior metric,
\begin{equation}
R=R_{0}\,x\,,\quad 0\leq x\leq 1\text{,}
\end{equation}
\begin{equation}
t=\frac{m\,T}{R_{0}}\,,  
\label{tT}
\end{equation}
where $x$ and $T$ are new coordinates. Then, the limit
$R_{0}\rightarrow 0$ can be taken safely and we obtain the metric of
the interior,
\begin{equation}
ds^{2}=-\left( 1+\frac{1}{n}-\frac{x^{n}}{n}\right) ^{-1}dT^{2}+m^{2}\left(
1+\frac{1}{n}-\frac{x^{n}}{n}\right) ^{2}\left( dx^{2}+x^{2}d\Omega
^{2}\right) .  \label{Tx}
\end{equation}
It is regular everywhere inside but incomplete for $x\leq 1$. It can
be extended at least up to a singular $x_{\mathrm{s}}$, given by
$x_{\mathrm{s} }=(n+1)^{1/n}>1$, but this singularity has nothing to
do with our original system. Now, it is seen from (\ref{tT}) that,
indeed, an infinite redshift occurs in the limit $R_{0}\rightarrow 0$
since finite intervals of $T$ correspond to infinitely growing
intervals of $t$. This mismatch in time scales gives a clear example
of why particles from the inside cannot penetrate to the outside.

We can observe one more important feature here. It is essential that
at $x=1$ (defining the boundary between dust and vacuum) the metric
(\ref{Tx}) has no horizon. Meanwhile, the outer metric represents an
extremal Reissner-Nordstr\"{o}m BH with the metric
\begin{equation}
ds^{2}=-\left( 1-\frac{m}{r}\right) ^{2}dt^{2}+\left( 1-\frac{m}{r}\right)
^{-2}dr^{2}+r^{2}d\Omega ^{2},  
\label{ern}
\end{equation}
and has a horizon at the boundary in this limit. Therefore, we cannot
match smoothly the two geometries: the surface $r=m$ (also given by
$x=1$) is timelike when seen from inside and is lightlike when seen
from outside. One may try to reconcile these two features by adopting
the original time coordinate $t$ inside as well. Then, in the limit
$R_{0}\rightarrow 0$ the interval along $r=m$ does indeed become null
from inside. However, this is achieved at the expense of the metric
becoming degenerate inside, since the term in $dt^{2}$ vanishes
everywhere in the inner region. Thus, in any case, spacetime as whole
exhibits singular, degenerate, features.

\subsubsection{Bonnor stars extended: continuous distribution of extremal
charged dust that asymptotes to the extremal Reissner-Nordstr\"{o}m geometry
\protect\cite{exdust}}

\label{extendedbonnorstars}

\textit{Generic properties:}

In \cite{exdust} Bonnor stars were modified, so that instead of having
a boundary where the charged extremal dust and the extremal
Reissner-Nordstr\"{o}m vacuum match, one has now a continuous,
extended, distribution of extremal charged dust which asymptotes to
the extremal Reissner-Nordstr\"{o}m geometry. This type of
distributions is specially interesting since it allows for cases where
there is a kind of hair when the QBH is forming, although these cases
are not going to be discussed here. Now, it is useful to rewrite the
metric (\ref{ab}) given in the Schwarzschild coordinate $r$, into an
isotropic form (\ref{dust0}), given in the coordinate $R$.

In \cite{exdust}, in these coordinates, the trial distribution is given by
the following form of the potential
\begin{equation}
\sqrt{B}=\frac{z}{z+q}\,,  
\label{dust1}
\end{equation}
where
\begin{equation}
z\equiv \sqrt{R^{2}+c^{2}}\,,  
\label{dust2}
\end{equation}
$c$ is a constant that can be chosen arbitrarily, and $q$ can be
thought of as the total charge, as we will see below. One can also
find that the potential $V$ defined in (\ref{defintionofV}) is given
by
\begin{equation}
\sqrt{V}=\frac{z^{3}+qc^{2}}{z^{2}(q+z)}\,.  
\label{dust3}
\end{equation}
Then, the metric can be rewritten as
\begin{equation}
ds^{2}=-\left( \frac{z}{z+q}\right) ^{2}\,dt^{2}+\frac{\left(
q+z\right) ^{2} }{z^{2}-c^{2}}\,dz^{2}+\frac{\left( z^{2}-c^{2}\right)
\left( z+q\right) ^{2} }{z^{2}}\,d\Omega ^{2}\,, 
\label{mz}
\end{equation}
valid for $z\geq c$. The density is then given by
\begin{equation}
\rho =\frac{3\,q\,c^{2}}{4\pi \,z^{2}\left( q+z\right) ^{3}}\,.
\label{density2}
\end{equation}
Then, one obtains that a quasihorizon forms at $r^{\ast }$, such that
for $ c\ll q$ one has $R=R^{\ast }\simeq q\left(
\frac{2c^{2}}{q^{2}}\right) ^{1/3} $. The explicit asymptotic behavior
near the quasihorizon reads,
\begin{equation}
\sqrt{B}=2^{1/3}\left( \frac{c}{q}\right)
^{2/3}+\frac{2}{3}\frac{(r-r_{\ast
})}{q}+\frac{2^{2/3}}{9c^{2/3}q^{4/3}}\left( r-r_{\ast }\right) ^{2}
...\,,
\label{asymptotic1}
\end{equation}
\begin{equation}
V=\varepsilon +\frac{2(r-r_{\ast })}{q^{2}}^{2}+...\,,  
\label{asymptotic2}
\end{equation}
where,
\begin{equation}
\varepsilon =\frac{9}{2^{4/3}}\left( \frac{c}{q}\right) ^{4/3}\,,
\label{asymptotic3}
\end{equation}
and in this limit, $r^{\ast}=q$.
So, near the formation of the QBH, for $c\rightarrow 0$, one finds there are
three characteristic regions. They are:

\vskip0.3cm (I) The inner core region $r\lesssim c$: Here it is convenient
to make the substitution $z=cy$ and take the limit $c\rightarrow 0$
afterward. Then, rescaling time as
\begin{equation}
t=\frac{q}{c}\,\tilde{t}\,,  \label{transf1_1}
\end{equation}
and making one more substitution
\begin{equation}
\cosh u\equiv y=\frac{z}{c}\,,  \label{transf1_2}
\end{equation}
one obtains
\begin{equation}
ds^{2}=q^{2}\left( -\cosh ^{2}u\,d\,\tilde{t}^{2}+du^{2}+\tanh
^{2}u\,d\Omega ^{2}\right) \,.  \label{m1}
\end{equation}
Thus the metric is everywhere regular. If one allows $u\rightarrow \infty $,
it becomes geodesically complete and asymptotically approaches the
Bertotti-Robinson metric \cite{br1,br2}.

\vskip0.3cm (II) The vicinity of the quasihorizon $r=r^{\ast }=q$: Then, it
is convenient to make the rescaling to the coordinates $T$ and $\eta $,
\begin{equation}
T=z^{\ast }\frac{t}{q}\,,  \label{definitionT}
\end{equation}
and
\begin{equation}
\eta =\frac{z}{z^{\ast }}\text{,}  \label{definitioneta}
\end{equation}
where $z^{\ast }=z(r^{\ast })$ and $\eta \leq 1$ corresponds to the inner
region. Then one can find by direct substitution that in the limit $
c\rightarrow 0$ the metric takes the form $ds^{2}=-\eta ^{2}\,dT^{2}+q^{2}(
\frac{d\eta }{\eta ^{2}}^{2}+d\Omega ^{2}),\label{definingmetric1}$ and
defining a new radial coordinate $l$ by $\eta =\exp \left( {\frac{l}{q}}
\right) $ with $l<0$, one has
\begin{equation}
ds^{2}=-\exp \left( \frac{2l}{q}\right) \,dT^{2}+dl^{2}+q^{2}d\Omega ^{2}\,.
\label{br}
\end{equation}
This metric is nothing else than the extremal version of the
Bertotti-Robinson metric \cite{br1,br2}. The region with $r\neq r^{\ast }$
is simply removed from the manifold. The coordinate $l$ can now be extended
into its full range, i.e., $-\infty <l<\infty $. As is known, the
Bertotti-Robinson spacetime is geodesically complete and, through yet
another coordinate transformation, can be cast into a form where the horizon
is absent (see, e.g., \cite{lap}). It is instructive to note that the
Bertotti-Robinson metric can be obtained also as an extremal limit of a
nonextremal Reissner-Nordstr\"{o}m spacetime. However, in that case the
resulting metric takes the form of the nonextremal version of
Bertotti-Robinson metric \cite{rn-br1,rn-br2}.
Note that, actually, regions I and II represent two different subregions of
the inner region inside the quasihorizon. If one makes the substitution $
y=\eta \frac{ z^{\ast }}{c}$, it becomes clear that (\ref{m1}) transforms to
(\ref{br}), provided $\eta \gg \frac{c}{z^{\ast }}\sim c^{1/3}$.

\vskip0.3cm (III) The region $r>r^{\ast }$: Here one can take the
limit in (\ref{mz}) directly and obtain the extremal
Reissner-Nordstr\"{o}m metric. As is known, the region $r>r^{\ast }$
represents only part of the extremal Reissner-Nordstr\"{o}m geometry
(\ref{ern}).

\vskip0.3cm We see, that from a formal viewpoint, the three different
spacetimes that arise from a single one, when the QBH forms, illustrate the
fact that the result of taking the appropriate limit depends strongly on how
the coordinates are involved in it \cite{geroch}. Indeed, for very small but
nonzero $c$ we have three distinct regions in the whole spacetime, a
spacetime that possesses no horizon. Each of those regions approaches the
corresponding form in its domain of validity: region (I) represents the
inner core region, region (II) gives the vicinity of the quasihorizon, and
region (III) corresponds to the outer solution. The spatial geometry for $r$
close to $r^{\ast }$ represents an extended throat on both sides of the
quasihorizon. The energy density $\rho (r^{\ast })\sim c^{2/3}\rightarrow 0$
\cite{exdust}. In the limit $c=0$, each of the three regions looks
incomplete in the original range of coordinates but can be made complete
after extension and proper continuation of coordinates into the whole
region. Similarly to the example (\ref{dust0})-(\ref{uext}), one can observe
from (\ref{br}) that the surface $r=r^{\ast }$ looks timelike from inside but
lightlike outside, so a smooth matching is impossible.

We also note that the Bertotti-Robinson spacetime can appear as a result of
two different limiting procedures, by taking a special portion of the
extremal Reissner-Nordstr\"{o}m metric and taking an appropriate limit, or
by taking a special limit of the QBH case under discussion. For further
details see Appendix \ref{a1}.

\vskip 0.5cm \textit{The total mass:}

It is instructive to compare the contribution of the inner and outer
regions to the total mass. Using the usual formulas for the energy
density and relationship between $r$ and $R$ one can obtain that the
proper mass $m_{ \mathrm{p}}=3\,q\,I$, where $I$ is given by,
$I=\int_{0}^{\infty }\frac{
dy\,y^{2}}{(1+y^{2})^{5/2}}=\frac{1}{3}$. So, $m_{\mathrm{p}}=\,q$.
Generically, for Majumdar-Papapetrou systems the proper mass is equal
to the electric charge. So $q$ has the meaning of total charge and
does not depend on $c$. From the calculation, one also finds that the
major contribution comes from the the inner region, $0\leq y\leq $
$y^{\ast }$, where $y^{\ast }\equiv \frac{R^{\ast }}{c}=2^{1/3}\left(
\frac{q}{c}\right) ^{1/3}\gg 1$.  The contribution from the outer
region is of the order $\frac{1}{2y^{\ast }}$ and becomes negligible
in the limit $c\rightarrow 0$. The same is true for the ADM
mass $m$. Thus, the competition from the two
factors, infinite proper volume (due to the extended throat) and
vanishing energy density, results in finite proper and ADM masses,
$m_{ \mathrm{p}}$ and $m$, respectively. The mass is concentrated
under the quasihorizon. Indeed, it is seen from (\ref{density2}) that
in the limit $ c\rightarrow 0$ the density $\rho \rightarrow 0$
everywhere except in the region of small $z\sim c$ near the origin.

\vskip0.5cm \textit{The curvature tensor and impenetrability:}

 From region I to II and vice versa: As the inner region I is geodesically
complete in the QBH limit and is at infinite proper distance from the
quasihorizon, there is really no question about penetrability from I to II
(which is adjacent to the quasihorizon) or III and vice versa. Indeed,
geodesics from region I can never reach regions II and III.

 From the inner region II to III (i.e., from the vicinity of the
quasihorizon to the outside) and vice versa: Near the quasihorizon
$z=z^{\ast }$, the components of the curvature tensor, following the
previous adopted nomenclature, are $K(r^{\ast
})=-\,\frac{1}{q^{2}}\,$, $\bar{K} (r^{\ast })=O(c^{2/3})\rightarrow
0\,$, $F(r^{\ast })=\frac{1}{q^{2}}\,$, $\bar{F} (r^{\ast })=0$. Thus,
in this sense the geometry is perfectly regular.  However, in the
free-falling frame, the quantity $\tilde{Z}$ is of the order
$c^{-2/3}$ and diverges. So, a particle cannot penetrate from the
outside to inside because infinite tidal forces appear, exactly in the
manner it was explained above while discussing the pure Bonnor
stars. In addition, the arguments presented previously for the pure
Bonnor stars, show that a particle with a finite energy measured with
respect to rescaled time $T$ of region II cannot penetrate from the
inner region to the outer one. As a result, regions II and III are
mutually impenetrable.

\vskip0.5cm \textit{Other considerations:}

Generalizing the approach of \cite{exdust}, we can notice that for a
continuous distribution of matter QBHs should always exist provided
\begin{equation} \rho (r_{\ast })\sim p_{r}(r_{\ast })\leq
O\,(\sqrt{\varepsilon })\,, \label{ror} \end{equation} where $p_{r}$
is the radial pressure, $\sqrt{B(r^{\ast})}\rightarrow 0$ and $
V=\varepsilon +a(r-r^{\ast })^{2}+...\,$, with $a$ a constant and $
\varepsilon \rightarrow 0$. These properties indeed hold for the
extremal dust solutions considered above \cite{exdust}. It follows
from Einstein equations that the metric function $\sqrt{B}$ obeys,
$\sqrt{B} =\sqrt{V} \exp\,(\psi )$, where $\psi =4\pi \int dr\,
\frac{r(\rho+p_{r} )}{V}\,$.  Then an elementary evaluation shows
that, on the quasihorizon, $\psi$ remains finite in this limit due to
the property (\ref{ror}). This entails that $\sqrt{B(r^{\ast})}\sim
\sqrt{\varepsilon }\rightarrow 0$. Since for QBHs, and in particular
for Majumdar-Papapetrou dust, one has $\frac{d\, \sqrt{B}}{dr}>0$ (see
Appendix \ref{a2}), this also means that $\sqrt{B} \rightarrow 0$ for
all $r<r^{\ast }$, and we return to the situation discussed
above. However, without knowing the details of the system, one cannot
state in advance whether or not the entire inner region will be
regular.

\subsection{Yang-Mills$-$Higgs matter and gravitational magnetic
monopoles
\protect\cite{lw1,lw2}}

The 't Hooft-Polyakov magnetic monopole, with a global magnetic
charge, is a solution of the Yang-Mills$-$Higgs system with no
gravity. When one couples gravitation, new important features
arise. This Eintein$-$Yang-Mills$-$Higgs system possesses regular
self-gravitating solutions for a range of parameters. In addition, for
a sufficiently massive monopole the system turns into an extremal
configuration. It was noted in \cite{lw1,lw2} that such an extremal
configuration is a QBH. Indeed, in those works it was coined for the
first time the word QBH, to distinguish such an object clearly from an
extremal BH. Such a magnetic QBH develops then an extremal
quasihorizon, with all the nontrivial matter fields inside it. For our
purposes here we note that the metric used for the gravitational
magnetic monopoles is of the type given in equation (\ref{ab}), and
that the numerical calculations carried out in \cite{lw1,lw2} show
that $\sqrt{B} \sim \varepsilon ^{q}$, where $q$ ranges between 0.7
and unity, and that $ \tilde{Z} \sim \varepsilon ^{-2q}$, where
$\tilde Z$, defined in the previous section, is a quantity related to
the tidal forces in a free-falling frame. In turn, this implies that
in a static frame the quantity $Z$ is regular, but in a free-falling
frame $\tilde Z$ diverges.  Thus, we have again the combination of a
perfectly regular geometry with a naked-type behavior inside the
entire inner region, as was observed in \cite{lw1,lw2}. The other
properties of QBHs discussed in the previous subsection follow through
a comparison between the properties of the Yang-Mills$-$Higgs system
with its gravitational magnetic monopoles and the corresponding QBHs
and the much simpler Majumdar-Papapetrou system with its Bonnor stars,
along the lines of \cite{jmp}.

\subsection{Vacuum with a surface layer: gluing between the extremal
Reissner-Nordstr\"om and other metrics}

\subsubsection{Gluing between the extremal Reissner-Nordstr\"om and
Bertotti-Robinson metrics \protect\cite{mod,n}}

\label{vacbert}

\vskip 0.2cm \textit{Generic properties:}

In \cite{mod,n} gluing between the extremal Reissner-Nordstr\"{o}m and
Bertotti-Robinson metrics \cite{br1,br2,lap,geroch,rn-br1,rn-br2} was
considered as a classical model of an elementary particle that looks
as a BH for an external observer but is regular inside. Let $m$ be the
ADM mass of such a BH, $r_{0}$ being the radius of gluing. For $r\geq
r_{0}$ we have the extremal Reissner-Nordstr\"{o}m metric (\ref{ern}),
with $B=(1-\frac{m}{r} )^{2}$, and for $r\leq r_{0}$ the metric has
the form (\ref{br}) with $ q=r_{0}$. Then, as is shown in \cite{mod},
the only nonvanishing component of the boundary surface stresses is
equal to $S_{0}^{0}=\frac{\sqrt{B(r_{0})} }{4\pi
r_{0}}=\frac{\varepsilon }{4\pi r_{0}^{2}}$, where $\varepsilon
=r_{0}-m$. For small but nonzero $\varepsilon $ we have the
configuration typical of a QBH: a static metric with the radius of the
inner region arbitrarily close to that of the horizon. In the limit
$\varepsilon \rightarrow 0$ the quantity $S_{0}^{0}\rightarrow 0$. For
an extremal Reissner-Nordstr\"{o}m BH the electric charge and the mass
obey $Q^{\mathrm{RN}}=m$. On the other hand, for the
Bertotti-Robinson metric (\ref{br}), one has $Q^{\mathrm{BR}}=q$, so
that in our case $Q^{\mathrm{BR}}=r_{0}$.  As a result, the shell
separating two regions carries the charge $Q^{\mathrm{ RN}
}-Q^{\mathrm{BR}}=-\varepsilon $ which also vanishes in the limit $
\varepsilon \rightarrow 0$. Thus, in the static coordinate frame, in
the quasihorizon limit $\varepsilon =0$ , one obtains that the surface
stresses and the surface charge (that appear due to the gluing process
between the two different metrics) vanish \cite{mod}, so that the
configuration becomes everywhere regular. For an outer observer, the
corresponding spacetime reveals itself as an extremal BH but it is
free of singularities inside (in contrast to the
Reissner-Nordstr\"{o}m metric) since the inner Reissner-Nordstr\"{o}m
core is replaced by the Bertotti-Robinson metric. One obtains a
self-sustained configuration having no singular sources, which is
balanced by its own forces without support from an external agent. In
this sense, it can be considered as a classical electromagnetic model
of an elementary particle, realizing Wheeler's idea of charge without
charge \cite{wheeler1}.

\vskip 0.1cm \vskip 0.5cm \textit{Tidal stresses and matter stresses:}

Now, let us see what happens in a freely falling frame. A free-falling
frame reveals some new nontrivial features of the composite spacetime
under discussion. Consider again the quantity $Z=\bar{F}-\bar{K}$,
with $\bar{K} =R_{\hat{0}\hat{\theta}}^{\hat{0}\hat{\theta}}$ and
$\bar{F}=R_{\hat{\theta} \hat{r}}^{\hat{\theta}\hat{r}}$, which we
introduced in Sec. III A while discussing some properties of extremal
charged dust. In the free-falling frame the quantity $\tilde{Z}$ is
given by $\tilde{Z}=Z\,(2\frac{E^{2}}{B} -1) $, where $E$ is the
energy of the particle and $\tilde{Z}$ is calculated in the
free-falling frame \cite{nk1,nk2} (see also \cite{vo,ssn,bm}). For the
Bertotti-Robinson metric one has $Z\equiv 0$, so that one obtains $
\tilde{Z} =0$. Therefore, one may wonder whether or not the naked
behavior typical of other examples of QBHs occurs in this case. As we
will show now, an analogue of naked behavior does indeed occur. Since
now $Z=0$ there is no naked behavior in the components of the Riemann
tensor inside the boundary surface (i.e. in the Bertotti-Robinson
region) but there is naked behavior in the components of the Ricci
tensor (i.e., in the components of stresses) on the boundary surface
itself. Let us see this in more detail. If one defines $Y\equiv
S_{1}^{1}-S_{0}^{0}$, then a local Lorentz boost leads to the
expression $\tilde{Y}=(2\frac{E^{2}}{B}-1)Y$ since $Y$ transforms like
$Z$, with the $\theta -\theta $ components being insensitive to
radial boosts. For the system under consideration, the only
nonvanishing component in the static frame is
$S_{0}^{0}=\frac{\varepsilon }{4\pi r_{0}^{2}}$ (see above). As a
result, in a free-falling frame $\tilde{Y}=\frac{1}{4\pi r_{0}^{2}}
\,\left( \varepsilon -2\,\frac{E^{2}r_{0}^{2}}{\varepsilon } \right)
\sim \varepsilon ^{-1}$, which clearly diverges in the QBH limit, $
\varepsilon \rightarrow 0$. Thus, we have displayed a remarkable
result: for $\varepsilon \neq 0$ the boundary stresses are finite and
nonzero, both in the static and free-falling frames. However, in the
limit $\varepsilon \rightarrow 0$, they disappear in the static frame,
but go unbounded in the free-falling one. In this sense, the situation
in the electrovacuum case is totally similar to that discussed above
for the extremal dust and non-Abelian gauge systems. The only
difference is that now the relevant quantities are not curvature
components but boundary stresses.

\subsubsection{Gluing between the extremal Reissner-Nordstr\"{o}m and
Minkowski metrics \protect\cite{vf}}

\label{vacmink}

An even simpler example can be invoked, where gluing between an inner
flat metric and an external extremal Reissner-Nordstr\"{o}m metric is
performed.  Such a construction was discussed in \cite{vf} as an
example of a classical model of an elementary particle (see also
\cite{jmp} and \cite{mod}).  Consider an external spacetime given by
equation (\ref{ern}) for $r\geq r_{0} $, and an inner spacetime given
by the Minkowski metric,
\begin{equation}
ds^{2}=-dT^{2}+dr^{2}+r^{2}d\Omega ^{2}  \label{mk}
\end{equation}
where $0<r\leq r_{0}$. On the border, the condition of matching both parts
of the spacetime leads to
\begin{equation}
t=\frac{Tr_{0}}{r_{0}-m}\text{,}  \label{tt}
\end{equation}
so that the time part of the metric (\ref{mk}) can be written as $
-\,dT^{2}=- \frac{(r_{0}-m)^{2}}{r_{0}^{2}}\,dt^{2}$. Then, if time
$t$ is used, the metric coefficient $g_{00}\rightarrow 0$ in the limit
$ r_{0}\rightarrow m$. This is the reason why this construction can be
considered as an example of a QBH. We again obtain an infinite
redshift due to the mismatch in time rescaling in equation
(\ref{tt}). Also, we cannot achieve the continuous matching if $T$ is
considered as a legitimate coordinate inside since the surface $r=m$
is timelike in the metric (\ref {mk}) but lightlike in the metric
(\ref{ern}). One may try to repair this by considering inside the same
time $t$ as outside. However, the term
$-\frac{(r_{0}-m)^{2}}{r_{0}^{2}} \,dt^{2}$ disappears in this limit
and the spacetime becomes degenerate. If one calculates the surface
stresses on the boundary, it turns out that $8\pi
S_{0}^{0}=-\frac{2}{m}\neq 0$ (all other components vanish)
\cite{mod}. Then, reasonings from the previous subsection
\ref{vacbert} apply and we obtain a naked behavior on the shell in the
limit under discussion for a radially infalling observer.

\section{A further property: quasi-black holes should be extremal}

\label{furtherprop}

In all examples considered above the horizon approached by the system
is extremal. One may ask, whether or not QBHs with nonextremal
horizons are possible. In \cite{df}, with the help of numerical
calculations, it was shown that, for some particular charge density
and energy density distributions, the boundary of a body with $q<m$
(where $q$ is the total charge and $m$ is the ADM mass) cannot
approach its own horizon, the system collapses before reaching
it. This result corroborates the Buchdahl limits, first worked out to
the Schwarzschild interior solution, as well as for perfect fluid
matter. In \cite{simingyunqiang} an interesting, although convoluted,
analytical proof generalizing the Buchdahl limits for charged perfect
fluid was given. Here we state an even more general theorem, without
resorting to the equation of state of the matter or other system's
details at all.

The statement we want to prove is \textquotedblleft a static regular
configuration cannot approach its own horizon arbitrarily closely if
the horizon is nonextremal.\textquotedblright\ The proof goes as
follows. By definition, a nonextremal horizon (NEH) implies that the
surface gravity $ \kappa $ is nonzero. Since $\kappa =\left(
\frac{d\sqrt{B}}{dl}\right)_{h}$, where the derivative is taken on the
horizon $h$, this condition gives $ \left( \frac{d\sqrt{B}}{dl}\right)
_{h}\neq 0$. We call this the NEH condition. Let conditions (a)-(c) of
Sec. \ref{def} be fulfilled, so that we have a QBH. Consider
separately two cases, namely, (1) $\sqrt{B(r)}$ has a continuous
derivative in relation to the proper length $l$, i.e., $\frac{d
\sqrt{B(r)}}{dl}$ is continuous, and (2) $\sqrt{B(r)}$ is merely
continuous, so that a surface layer is allowed.

\begin{description}
\item[(1)] When $\sqrt{B(r)}$ has a continuous derivative in relation
to $l$ , i.e., $\frac{d\sqrt{B(r)}}{dl}$ is continuous, then also
$\frac{d\sqrt{ B(r) }}{dr}$ is continuous. Thus we are assuming that
$\sqrt{B(r)}$ is of class C$^{1}$. Now we will show that the NEH
condition and condition (c) are mutually inconsistent. Recall that
condition (c) states that in the limit $ \varepsilon \rightarrow 0$
the metric coefficient $B\rightarrow 0$ for all $ r\leq r^{\ast}$. Let
us exploit the following simple lemma, which we will prove
shortly. Assume $\sqrt{B}$ is any function such that the condition (c)
is satisfied, and further assume (d) $\sqrt{B}>0$, for $\varepsilon
\neq 0$ and $r\leq r^{\ast}$. Then in the limit $\varepsilon
\rightarrow 0$ one cannot get $\frac{d\sqrt{B(r)}}{dl}\neq 0$, as one
should for a nonextremal BH. Now we prove this lemma. Let us suppose,
for a moment, that $\frac{d \sqrt{B}}{dr}\rightarrow a_0\neq 0$ at
some $r_{1}$ where $0\leq r_{1}<r^{\ast }$. Using a Taylor expansion,
we can write $\sqrt{ B} =a_0\,(r-r_{1})+...$ in the vicinity of
$r_{1}$ for sufficiently small $ \varepsilon$, with $a_0$ a
constant. For $a_0>0$ we have that $\sqrt{B}<0$ for $r<r_{1}$ in
contradiction with condition (d). As well, for $a_0<0$ we have that
$\sqrt{B}<0$ for $r>r_{1}$ in contradiction with condition (d). So the
only possibility is $\frac{d\sqrt{B}}{dr}\rightarrow a_0=0$. As, by
assumption, the derivative $\frac{d\sqrt{B}}{dr}$ is continuous, we
can extend this line of reasoning to some vicinity $(r^{\ast}-\delta
$, $r^{\ast }+\delta )$ of the boundary point $r^{\ast }$, take
advantage of the Taylor series again, by the same reasoning obtain
that $\left( \frac{d\sqrt{B}}{dr} \right) _{r=r^{\ast }}\rightarrow
0$, and so, $\left( \frac{d\sqrt{B}}{dl} \right) _h\rightarrow 0$ as
well.
\item[(2)] When $\sqrt{B(r)}$ is merely continuous, one is relaxing
the condition of the continuity of the first derivative and thus
allowing the existence of a surface layer. We will see now that it
does no good. In this case we would have a deltalike term in the
stress-energy tensor $\tilde{T} _{\mu }^{\nu }$ and a nonzero Lanczos
tensor $S_{\mu }^{\nu }=\int dl\, \tilde{T}_{\mu}^{\nu}$, where the
integral is to be performed across the boundary. There is only one
independent spatial component of the tangential stresses, namely
$S_{2}^{2}$. This is given by, $8\pi\,S_{2}^{2}= \frac{ \left(
r\right)_{+}^{\,\prime}-\left( r\right)_{-}^{\,\prime}}{r^{\ast }} +
\frac{\left(\sqrt{B}\right)_{+}^{\,\,\prime } - \left(\sqrt{B}
\right)_{-}^{\,\,\prime}}{\sqrt{B}}\,, \label{bs}$ where the $+$ and
$-$ signs refer to the outer and inner regions, respectively, and a
prime denotes a derivative with respect to the proper distance
$l$. The first term is finite and is equal to zero, if we do not put a
finite mass on the surface $r=r^{\ast }$. However, the second term
diverges since the numerator is finite whereas the denominator tends
to zero. Thus, the boundary stresses become infinite and the
configuration becomes strongly singular.  
\end{description}
Thus, we see that in case (1) the condition of nonextremality cannot be
fulfilled, and in case (2) the condition of regularity fails. The proof of
our statement is completed, there are no nonextremal QBHs.

On the other hand, if the surface gravity $\kappa =0$, i.e., the QBH is
extremal, the above arguments do not work since $\frac{d\sqrt{B}}{dl}
\rightarrow 0$ from both sides of $r^{\ast }$. As a result, in case (1)
there is no contradiction between conditions NEH and (c), and in case (2) $
S_{2}^{2}$ can be finite. So QBHs can only be extremal.

\section{Regular versus singular behavior and unattainability of the
quasi-black hole limit}

\label{regular}

Upon careful inspection, one finds that in QBHs divergencies on the
Kretschmann scalar do not occur. However, the finiteness of this
quantity is not the only criterion for regular or singular
classification of a spacetime. One example is the behavior of naked
BHs. Indeed, in some special frames the Riemann tensor diverges near
the horizon of these naked BHs and these divergences can be related to
nonscalar polynomial curvature singularities discussed in \cite{es}.

In the present work we have encountered a rather unusual entanglement
of regular and singular features in QBHs. From the viewpoint of an
external observer who uses time measured by clocks at infinity, an
inner region looks like a degenerate spacetime with the component of
the metric $ g_{00}\rightarrow 0$ everywhere. Yet, this singular
feature has nothing to do with the behavior of the Riemann tensor. Its
components in an orthonormal static frame are finite there, and the
Kretschmann scalar is also well behaved. The most obvious
manifestation of this property is the example discussed in Section
\ref{vacmink} where the inner spacetime is flat, nonetheless it
exhibits singular features! If one tries to remove the degeneracy of
the inner spacetime by rescaling the time coordinate, another
difficulty arises: the spacetime ceases to be continuous since the
surface is lightlike from the viewpoint of an outer observer but is
timelike from the viewpoint of an inner one. To put it in another way:
one can easily achieve the validity of the matching conditions on a
timelike surface, but if this surface tends to a null surface, at
least from one side, the procedure ceases to be well-defined and this
gives rise to a number of unusual properties. Another singular feature
consists in the impossibility to penetrate from the inside to the
outside and vice versa. In this sense, geodesics cannot be extended
across the border between different regions, in spite of the fact that
each of them, taken by itself, can be extended. For instance, the
Minkowski spacetime in Section \ref{vacmink} is obviously extendable
but this extension has nothing to do with the problem under discussion
in which the outer spacetime should be the extremal
Reissner-Nordstr\"{o}m BH. The fact that observers in different
regions disagree about the border's nature, whether it is timelike or
null, can be considered as one of the manifestations of the mutual
impenetrability.  Actually, it shows that one deals with two separate
spacetimes. It turns out that there is some kind of complementary
relationship between the inner and outer regions and between their
regular and singular properties. If an observer is situated inside, he
will say that the geometry is perfectly regular there but becomes
singular on the border and beyond, so that he is unable to penetrate
to outside. The outer observer, on the contrary, will say that it is
his region which is regular (excepting the border) and finds he cannot
penetrate into the inner singular region. All this forces us to
conclude that the spacetime of a QBH as a whole may be singular in
spite of the fact that the Kretschmann scalar diverges nowhere.

This discussion helps to elucidate an important additional question,
of whether or not the QBH limit (whose properties we have discussed in
detail) is attainable in some real physical process. For comparison,
in the Reissner-Nordstr\"{o}m geometry, taking formally the limit
$q\rightarrow m$, one can obtain the extremal Reissner-Nordstr\"{o}m
BH from the nonextremal one but, according to the third law of BH
thermodynamics \cite{wald}, this cannot be accomplished in any real
process for a finite number of steps.  Furthermore, if the cosmic
censorship conjecture is valid, one cannot convert the BH state with
$q\leq m$ into a naked singularity by increasing the charge to
$q>m$. What is said above about singular features in the QBHs
properties leads to the conclusion that the corresponding limiting
state is unattainable physically from any close regular
configuration. More precisely, the state which is obtained by the
mathematical procedure of taking the QBH limit can be approached as
closely as one likes. However, if we assume that regular
configurations cannot be turned into singular ones, the QBH limit
cannot be attained by gradually changing an initial regular
configurations to a singular one. A usual horizon hides singularities
beyond it, but its analogue, the quasihorizon, in a sense, brings
about certain singular features into the system. If these singular
features cannot arise by physical processes, as we have argued, this
means that we are faced with a somewhat unusual counterpart of the
cosmic censorship. On the face of what has been said, it seems that
QBHs should extend the taxonomy, not only of relativistic objects, but
also of singularity types in general relativity.

It is also worth remarking that in some cases the limiting
configuration may turn out to be geodesically complete and regular
like the manifold given by equation (\ref{m1}), obtained from the
inner core region (i.e., region I), in Section
\ref{extremaldustbonnor}. In this case, nothing prevents one from
taking the limit $c=0$ in which, equation (\ref{m1}) arises from
equation (\ref{mz}). In addition, the proper distance to the
quasihorizon tends to infinity in this limit. Thus, it seems that the
limit can be attainable in some regions and unattainable in others,
which is one more unusual feature of QBHs.

Summing up, configurations that approach as close as one likes a QBH
state can be easily achieved, and in this sense, QBHs may have real
physical significance. But whether a QBH state can be attained in
nature, through such a process, or perhaps emerge via some quantum
process, is a thorny issue that certainly needs further investigation.

\section{Conclusions}

\label{conc}

The present work unifies in the same QBH context seemingly different
systems like those considered in 
\cite{bonnor1964,bonnor1975,bonnor1999,kleberlemoszanchin,jmp,exdust} 
on Bonnor stars, in \cite{lw1,lw2} on magnetic monopoles, and in
\cite{mod}-\cite {wheeler1} on glued vacua. The properties of QBHs
were worked out in some detail. It is then clear, that for an external
static observer, a BH and a QBH look similar. Nevertheless, their
inner nature is different. First, not only the outer original region
is inaccessible for the inner observer, like in the BH case, but also
vice versa, which has no analogue in the BH case.  Second, while for
BHs the separation of different regions is of pure causal nature, in
the QBH it is dynamic rather than purely causal. The reasons for no
penetration from one region to another are quite different, namely
rescaling of time, and infinite tidal forces or infinite surface
stresses, i.e., naked behavior. In addition, as far as the naked
behavior is concerned, it is also worth noting for comparison that in
all examples considered in \cite{nk1,nk2,vo,ssn,bm} the curvature
components in the free-falling frame are enhanced with respect to the
static value but remain finite, whereas for QBHs those components
diverge. Thus, if a system is able to withstand gravity forces up to a
state which is arbitrarily close to an extremal BH and not collapse,
its inner properties, the QBH properties, are qualitatively distinct
from those of a corresponding extremal BH. However, for a distant
observer to distinguish between a QBH and an extremal BH might be
virtually impossible.

As a last remark, we note that in the above considerations, we tacitly
implied that the areal radius increases monotonically with the proper
distance. Meanwhile, we can try to glue two copies of the spacetime in
the spirit of cut and paste technique used in physics of wormholes
\cite {bookvisser} with the increasing and decreasing branches of the
function $ r(l)$ and, afterward, take the QBH limit. For example, one
can use the extended Bonnor star distributions described above. The
corresponding limit possesses interesting properties that, however,
needs a separate discussion.  In \cite{damoursoloduk} a special type
of wormhole was considered.  Interestingly enough, this wormhole can
be considered as a system with properties somehow similar to those of
a QBH, in the sense that it is connected with the threshold of the
formation of a horizon, in this case nonextremal, from a wormhole
configuration. Detailed comparison of the two approaches, based on
near-extremal and nonextremal wormhole configurations, and properties
of the corresponding spacetimes will be done elsewhere.

\begin{acknowledgments}
\noindent JPSL thanks Funda\c{c}\~{a}o para a Ci\^{e}ncia e Tecnologia (FCT)
through project POCI/FP/63943/2005 for financial support. OZ thanks Centro
Multidisciplinar de Astrof\'{\i}sica - CENTRA for hospitality and a
stimulating atmosphere.
\end{acknowledgments}

\newpage
\appendix

\section{Obtaining the Bertotti-Robinson spacetime as a limit of different
metrics}

\label{a1}

It is instructive to note that the Bertotti-Robinson spacetime 
\cite{br1,br2,lap,geroch,rn-br1,rn-br2} can appear as a result of two
different limiting procedures.

(1) First, starting with the extremal Reissner-Nordstr\"{o}m metric
one obtains the Bertotti-Robinson metric by means of a well known
limiting procedure \cite{geroch}. Indeed, in the extremal
Reissner-Nordstr\"{o}m case one can make the transformation, from the
usual Schwarzschild coordinate $r$ to the proper radial coordinate
$l$, given by, $r=q+\lambda\exp\left( \frac{ l } {q}\right) $, and
from $t$ again to $T$ given by, $t=\frac{qT}{\lambda}$ , where
$\lambda$ is a parameter, and take the limit $\lambda\rightarrow0$.
Then the metric takes the form (\ref{br}). In the course of this
limiting transition the metric coefficient
$g_{00}^{\mathrm{\mathrm{RN}}}$ of the original extremal
Reissner-Nordstr\"{o}m metric tends to zero just due to taking the
limit in the coordinate space: since $\lambda\rightarrow0$ we have
$r\rightarrow q$ and $g_{00}^{\mathrm{\mathrm{RN}}}(r)\rightarrow
g_{00}^{\mathrm{\mathrm{RN}} }(q)=0$. In the resulting
Bertotti-Robinson manifold (\ref{br}) the coefficient
$g_{00}^{\mathrm{\mathrm{BR}}}\neq0$, due to the factor $\lambda^{-2}$
which compensates $\lambda^{2}$ in $ g_{00}^{
\mathrm{\mathrm{RN}}}$. In doing so, the horizon of the original
Reissner-Nordstr\"{o}m metric ($r=q$) maps into the horizon of the
Bertotti-Robinson metric ($l=-\infty$).

(2) Second, in the QBH case discussed in section
\ref{extendedbonnorstars}, the reason why $g_{00}\rightarrow0$ comes
from taking a special limit in the space of parameters: we have
$g_{00}^{\mathrm{QBH}}=g_{00}^{\mathrm{QBH} }(r,c)$ and
$g_{00}^{\mathrm{QBH}}(r^{\ast},c)\neq0$, for $c\neq0$. But $ \lim
_{{c}\rightarrow0}g_{00}^{\mathrm{QBH}}(r^{\ast},{c})=0$ where
$r^{\ast} $ corresponds to the quasihorizon \cite{exdust}. In doing
so, the quasihorizon $r=r^{\ast}$ corresponds to $l=0$, i.e., $\eta=1$
in (\ref {definitioneta}). Then, it is seen from (\ref{br}) that
$g_{00}^{\mathrm{\ \mathrm{BR}}}\neq0$ at $\eta=1$ and, thus, this
value of $\eta$ does not correspond to the horizon of the
Bertotti-Robinson metric. The horizon of the metric (\ref{br}) lies at
$l=-\infty$ where $g_{00}^{\mathrm{\mathrm{BR}} }\rightarrow0$. In
other words, the quasihorizon of the original metric ( \ref{mz}) does
not map onto the horizon of the Bertotti-Robinson obtained from it
through the limiting procedure. Instead, the transformation (\ref
{definitioneta}) in the limit $c\rightarrow0$ maps the origin $r=0$ of
(\ref{mz}) onto the horizon of the metric (\ref{br}) (which does not
posses a origin at all) since in this limit $r^{\ast}\sqrt{B}\sim
c^{2/3}$ and $ \eta\sim c^{1/3}\rightarrow0$ (see \cite{exdust}).

Thus, we see that although in region II our metric takes the
Bertotti-Robinson form, it cannot be considered as a trivial
consequence of the known limiting procedure from the extremal
Reissner-Nordstr\"{o}m metric. Cases (1) and (2) are different and map
the horizon, and the quasihorizon, of the original manifold in
different manners.

\section{Proof that $\frac{d\protect\sqrt{B}}{dr}\geq 0$ for quasi-black
holes}

\label{a2}

One general property of QBHs is that the metric function $\sqrt{B}$,
for the systems under discussion, obeys the condition
\begin{equation}
\frac{d\sqrt{B}}{dr}\geq 0\text{.}  \label{b'}
\end{equation}
Indeed, assuming that the Einstein equations are satisfied, one has
\begin{equation}
\frac{1}{\sqrt{B}} \frac{d\sqrt{B}}{dr}=\frac{m+4\pi p_{r}r^{3}}{r(r-2m)}
\text{,}  
\label{b+}
\end{equation}
where $m(r)$ is the total gravitational mass enclosed inside the
radius $r$, and $p_{r}$ is the total radial pressure, arising from all
the fields and matter that may be present. For regular matter
configuration there are no horizon, so the denominator is positive. In
addition, the numerator is positive for systems with $m(r)+4\pi
p_{r}r^{3}>0$. The known Majumdar-Papapetrou exact solutions show that
(\ref{b+}) holds for these systems
\cite{bonnor1964,bonnor1975,bonnor1999,kleberlemoszanchin,jmp,exdust}. For
the self-gravitating monopole its validity is clearly seen from
numeric calculations \cite{lw1,lw2}. For the composite vacuum systems
studied here \cite{mod,n}, composed of Reissner-Nordstr\"om and
Bertotti-Robinson geometries, the situation is more tricky, as the
coordinate $r$ becomes degenerate, but the positivity of (\ref{b+}) is
guaranteed upon a suitable redefinition of distance.


\newpage

\begin{thebibliography}{99}


\bibitem{maj} S. D. Majumdar, Phys. Rev. \textbf{72}, 390 (1947).

\bibitem{pt} A. Papapetrou, Proc. Roy. Irish Acad. A \textbf{51}, 191
(1947).

\bibitem{bonnor1964} W. B. Bonnor, Nature \textbf{204}, 868 (1964).

\bibitem{bonnor1975} W. Bonnor and S. B. P. Wickramasuriya, Mon. Not. R.
astr. Soc. \textbf{170}, 643 (1975).

\bibitem{bonnor1999} W. B. Bonnor, Class. Quant. Grav. \textbf{16}, 4125
(1999).

\bibitem{kleberlemoszanchin} A. Kleber, J. P. S. Lemos and V. T. Zanchin,
Grav. Cosmol. \textbf{11}, 269 (2005).

\bibitem{jmp} J. P. S. Lemos and V. T. Zanchin, J. Math. Phys. \textbf{47},
042504 (2006).

\bibitem{exdust} J. P. S. Lemos and E. Weinberg, Phys. Rev. D \textbf{69},
104004 (2004).

\bibitem{lw1} A. Lue and E. J. Weinberg, Phys. Rev. D \textbf{60}, 084025
(1999).

\bibitem{lw2} A. Lue and E. Weinberg, Phys. Rev. D \textbf{61}, 124003
(2000).

\bibitem{mod} O. B. Zaslavskii, Phys. Rev. D \textbf{70}, 104017
(2004).

\bibitem{n} O. B. Zaslavskii, Phys. Lett. B \textbf{634}, 111 (2006).

\bibitem{br1} B. Bertotti, Phys. Rev. \textbf{116}, 1331 (1959).

\bibitem{br2} I. Robinson, Bull. Acad. Pol. Sci. \textbf{7}, 351
(1959).

\bibitem{lap} A. S. Lapedes, Phys. Rev. D \textbf{17}, 2556 (1978).

\bibitem{geroch} R. Geroch, Commun. Math. Phys. \textbf{13}, 180
(1969).

\bibitem{rn-br1} O. B. Zaslavskii, Phys. Rev. Letters \textbf{76},
2211 (1996).

\bibitem{rn-br2} O. B. Zaslavskii, Phys. Rev. D \textbf{56}, 2188
(1997).

\bibitem{vf} A. V. Vilenkin and P. I. Fomin, Nuovo Cim. A
\textbf{\textbf{45}}, 59 (1978).

\bibitem{wheeler1} C. W. Misner, K. S. Thorne and J. A. Wheeler,
\textit{Gravitation}, (Freeman, San Francisco 1973), p. 1200.

\bibitem{nk1} G. T. Horowitz and S. F. Ross, Phys. Rev. D \textbf{56},
2180 (1997).

\bibitem{nk2} G. T. Horowitz and S. F. Ross, Phys. Rev. D \textbf{57},
1098 (1998).

\bibitem{vo} V. Pravda and O. B. Zaslavskii, Class. Quant. Grav.
\textbf{22}, 5053 (2005).

\bibitem{ssn} O. B. Zaslavskii, Phys. Rev. D. to appear (2007),
arXiv:0706.2727.

\bibitem{bm} I. S. Booth and R. B. Mann, Phys. Rev. D \textbf{59},
064021 (1999).

\bibitem{es} G. F. R. Ellis and B. G. Schmidt,
Gen. Rel. Grav. \textbf{8}, 915 (1977).

\bibitem{wald} R. M. Wald, Phys. Rev. D \textbf{56}, 6467 (1997).

\bibitem{buchdahl} H. A. Buchdahl Phys. Rev. \textbf{116}, 1027
(1959).

\bibitem{df} F. de Felice, L. Siming and Y. Yunqiang, Class. Quantum
Grav.  \textbf{16}, 2669 (1999).

\bibitem{simingyunqiang} Y. Yunqiang and L. Siming, Comm. Theor. Phys.
\textbf{33}, 571 (2000); arXiv:gr-qc/9905099.

\bibitem{thorne} K. S. Thorne, \textit{in Magic without Magic},
ed. J. R.  Klauder, (Freeman and Company, 1972), p. 231.

\bibitem{bonnor} W. B. Bonnor, Phys. Lett. A \textbf{102}, 347 (1984).

\bibitem{wheeler2} R. Ruffini and J. A. Wheeler, Phys. Today
\textbf{24}, 30 (1971).

\bibitem{bekenstein} J. D. Bekenstein, arXiv:gr-qc/9605059.

\bibitem{bardeen} J. M. Bardeen, \textit{in Proceedings of GR5},
Tiflis, USSR, (unpublished, 1968).

\bibitem{beato} E. Ay\'{o}n-Beato and A. Garc\'{\i}a, Phys. Lett. B
\textbf{493}, 149 (2000).

\bibitem{borde} A. Borde, Phys. Rev. D \textbf{55}, 7615 (1997).

\bibitem{bookvisser} M. Visser, \textit{Lorentzian Wormholes: From
Einstein to Hawking }(AIP Press, New York, 1995).

\bibitem{damoursoloduk} T. Damour and S. N. Solodukhin,
arXiv:0704.2667.

\bibitem{lemosreview} J. P. S. Lemos, \textit{in Proceedings of the
Fifth International Workshop on New Worlds in Astroparticle Physics},
eds. Ana M.  Mour\~ao et al (World Scientific, 2006), p. 71;
arXiv:gr-qc/0507101.

\bibitem{wl} A. Lue and E. Weinberg, Gen. Rel. Grav. \textbf{32}, 2113
(2000).




\end{thebibliography}
\end{document}